\documentclass[10pt,conference]{IEEEtran}

\usepackage[ruled,vlined]{algorithm2e}
\usepackage{epsfig}
\usepackage{url}
\usepackage{cite}
\usepackage{subfigure}
\usepackage{authblk}
\usepackage[skip=5pt]{caption}
\usepackage{color}
\IEEEoverridecommandlockouts

\providecommand{\eat}[1]{}

\begin{document}
\title{Demo Abstract: CDMA-based IoT Services with Shared Band Operation of LTE in 5G}

\author[1]{Siddarth Mathur}
\author[2]{Shweta S. Sagari}
\author[2]{Syed Obaid Amin}
\author[2]{Ravishankar Ravindran}
\author[3]{Dola Saha}
\author[1]{Ivan Seskar}
\author[1]{\\Dipankar Raychaudhuri}
\author[2]{Guoqiang Wang}
\affil[1]{WINLAB, Rutgers University, NJ, 
{\em\{siddarthmathur, seskar, ray\}@winlab.rutgers.edu}}
\affil[2]{Huawei Research Center, CA, {\em\{shweta.sagari, obaid.amin, ravi.ravindran, gq.wang\}@huawei.com}}
\affil[3]{University at Albany, SUNY, NY, {\em dsaha@albany.edu}}

\eat{
\author[1]{A1}
\author[1]{A2}
\author[2]{A3}
\author[3]{A4}
\affil[1]{Affiliation1, {\em\{a1, a2\}@email.com}}
\affil[2]{Affiliation2, {\em  a3@email.com}}
\affil[3]{Affiliation2, {\em a4@email.com}}
}

\maketitle
\thispagestyle{empty}
\let\VERBATIM\verbatim
\def\verbatim{%
\def\verbatim@font{\small\ttfamily}%
\VERBATIM}

\begin{abstract}

With the vision of deployment of massive Internet-of-Things (IoTs) in 5G
network, existing 4G network and protocols are inefficient to handle sporadic
IoT traffic with requirements of low-latency, low control overhead and low
power. To suffice these requirements, we propose a design of a PHY/MAC layer
using Software Defined Radios (SDRs) that is backward compatible with existing
OFDM based LTE protocols and supports CDMA based transmissions for low
power IoT devices as well. 
This demo shows our implemented system based on that design and the viability of the proposal under different network scenarios. 
\eat{As a proof-of-concept, we implemented the evaluate the proposed design under different scenarios
underlay CDMA-based IoT design
for scenarios of multi-IoT links connected to a Base Station (or eNodeB) in presence of an LTE link.}
\end{abstract}


\section{Introduction}
With exponential growth of IoT devices~\cite{CiscoVNI_Feb16}, the 5G network will experience a variety of traffic patterns not prevalent in earlier 4G systems.
IoT devices often transmits short sporadic messages, which are not well suited to the high data traffic and connection-oriented modes associated with legacy 3GPP networks resulting in high service latency and excessive control overhead. 
In order to access the network, a User Equipment (UE) has to follow attachment, authentication and bearer establishment procedure which account up to $30\%$ of control plane signaling overhead. 
Furthermore, a UE goes into the Idle state if it has been inactive for more than 10 seconds (applicable for many targeted IoT applications) and the UE needs re-establish the bearer for the next transmission. Latency for the Idle to connected state is $\sim 60$ ms\cite{Singhal2010_latencyAna}.
With the dense deployment of IoT devices, current 4G network will be extensively overwhelmed by the surge in both traffic and control plane signaling load.
The 5G network needs the design provision to accommodate heterogeneous IoT applications at very high scale with low latency and low control overhead across both the radio access network and core network. 

The goal is to operate in the same band as current LTE, thus not requiring any separate channel allocation, 
and is backward compatible with the 4G network\cite{Mathur2016_crossLayer}.
This motivates us to propose CDMA-based low power IoT  transmission for the shared band operation of IoT and LTE devices.
With the proposed CDMA-based cross-layer MAC and Physical layer solution, we achieve low latency, short-message and long range communication required for low-power IoT devices.
The details of proposed CDMA-based IoT protocol, hardware setup and evaluation scenarios are given in following sections.

\section{Proposed Design}
\label{sec:sysDes}
\begin{figure}[t]
\begin{center}
\vspace{0.2in} \hspace{0.35in}
\epsfig{figure= 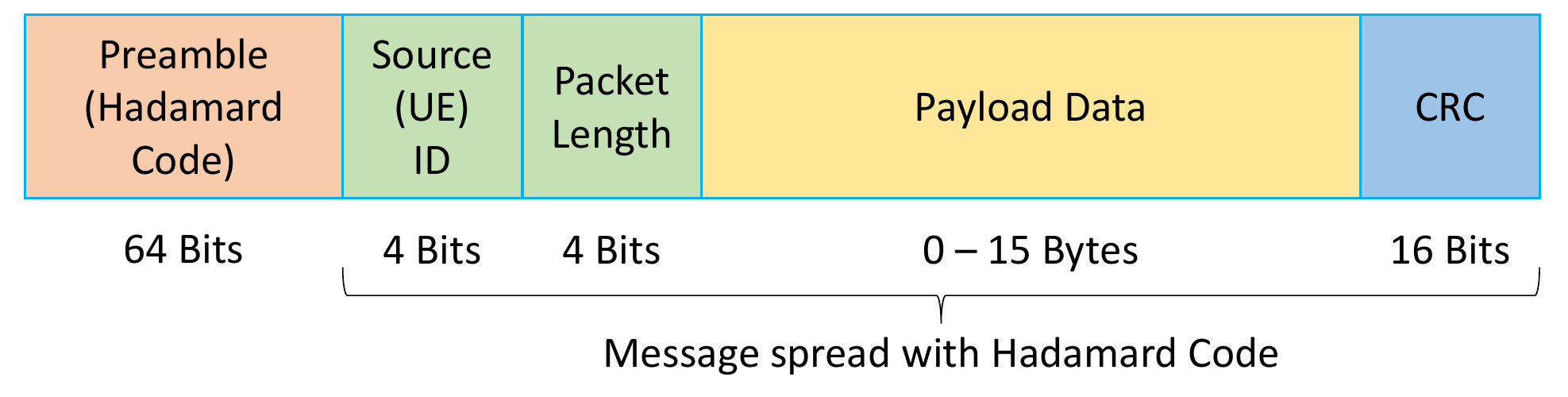,width=3.45in}
\end{center}
\caption{Packet format of CDMA-based IoT data transmission}
\label{fig:ProtoSynTax}
\vspace{-0.7cm}
\end{figure}
The main objective at MAC layer is to schedule IoT traffic with the minimum wait time to access the channel. At physical layer, we need to identify the optimal waveform for the\eat{an underlay} transmission operating at low signal-to-interference-plus-noise ratio (SINR) and bit-error-rate (BER) and without causing significant interference to the overlay LTE transmission. Considering these requirements, the CDMA underlay transmission becomes the favorable choice for IoT services due to (1) it does not require any UE-eNB handshaking (RACH+RRC) contrary to earlier efforts made in the community\cite{Schaich2014_waveform}, (2) Asynchronous CDMA transmission enables decentralized spectrum access at IoT UEs which reduces wait time to get assigned resources from eNB, (3) CDMA IoT transmission can reject narrow band OFDMA LTE interference without causing significant interference to LTE as well with its low-power transmission, and (4) CDMA-based IoT transmission utilizes the legacy CDMA support available at cellular network. One of the major challenges in CDMA transmission with LTE-overlay is interference cancellation at the receiver (here, eNB). Therefore, self-interference cancellation is introduced at eNB in full duplex mode where eNB cancels its own transmitted signal from the received signal to reduce the interference.

Fig.~\ref{fig:ProtoSynTax} shows a packet format instance for the uplink IoT data transmission. The CDMA MAC frame consists of source addresses, payload and cyclic Redundancy Check (CRC) to check errors in frame. This example frame format allows to send variable size payload of 0-15 Bytes which is typical IoT traffic profile emanating from sensing devices\cite{LPWA2016}. The MAC frame is spreaded/despreaded with the 64-bit Walsh-Hadamard code where separate codes are assigned to different UEs. Data is modulated/demodulated using BPSK to be able to operate on low-SINR. Packet preamble contains a unique Walsh-Hadamard code which is used to detect the start of the packet at the receiver by correlating signal with the known code. The CDMA channel uses 1 MHz bandwidth with data rate $\sim$10 - 15 Kbps. The proposed design parameters are consistent with existing low power wide area network IoT technologies\cite{LPWA2016} and also compatible with coexistence operation of LTE.

\section{System Implementation}
\begin{figure}[t]
\begin{center}
\vspace{0.2in} \hspace{0.35in}
\epsfig{figure= 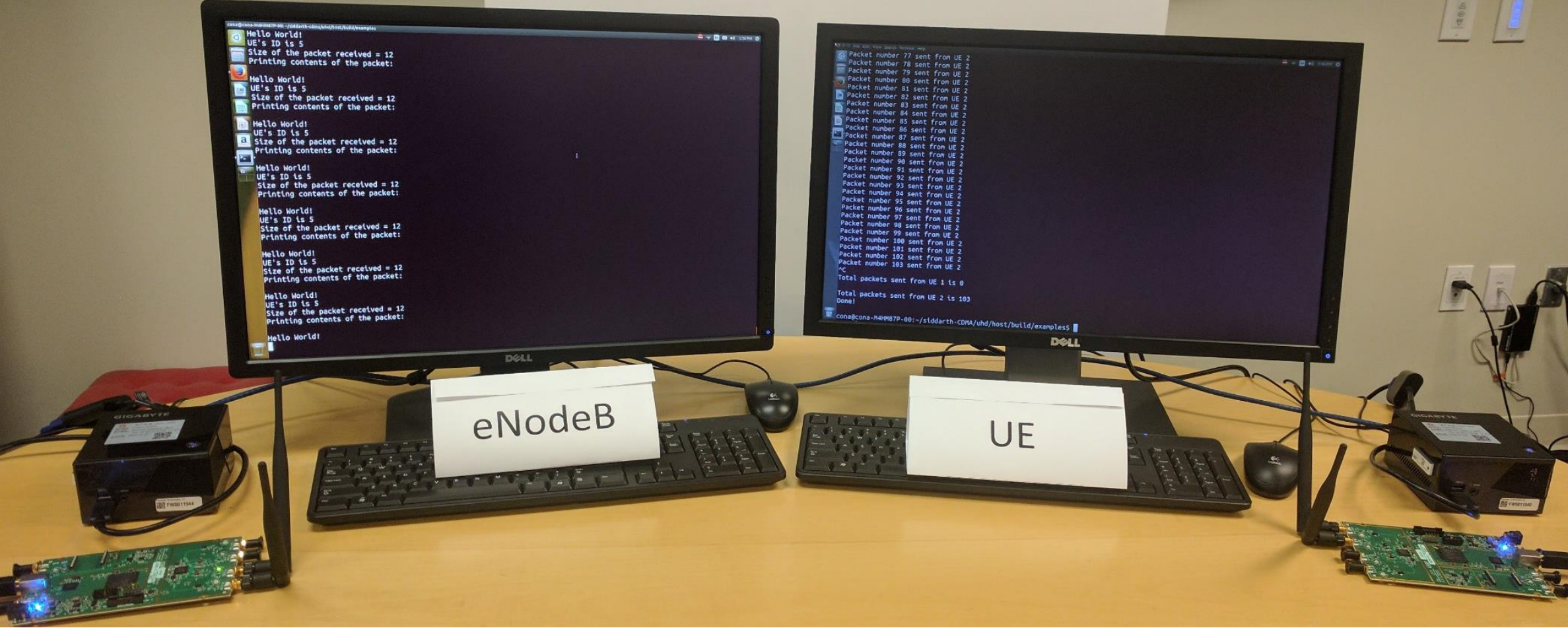,width=3.3in}
\end{center}
\caption{Demo setup of a CDMA-based IoT transmission}
\label{fig:demo_setup}
\vspace{-0.6cm}
\end{figure}
CDMA-based IoT transmission is prototyped using a software-defined radio (SDR) platform using GNU radio and  Universal Software Radio Peripheral (USRP) (see Fig.\ref{fig:demo_setup}). USRP Hardware Drivers (UHD) are used to transmit/receive samples to/from the USRP. The CDMA transmitter and receiver code is developed on top of UHD in C++ and C, respectively, to process and decode CDMA packets in real time. Intel(R) Core i7 4th Generation CPU ($@$3.60 GHz) machines are used in performance mode. Developing CDMA receiver is particularly challenging to detect CDMA packet considering random wireless channel and critical time constraints of real-time signal processing. For example, to detect beginning of the packet, we perform cross-correlation of a known 64-bit preamble and 10,000 received samples (equivalent to size of a packet) at one instance. Each instance takes processing time of 160 $\mu$s including 100 $\mu$s for reading samples at sampling rate of 1 MSps and 60 $\mu$s for running correlation function. We choose sampling rate 1 MSps to avoid overflowing of samples at receiver which causes due to higher sampling rate. This parameter also restrict the maximum achievable data rate for the CDMA transmission. 
Furthermore, a packet is detected if the peak value of the cross-correlation output is greater than certain threshold which is a function of SINR. Here the choice of the threshold becomes critical. If it very low, then there is significant false packets detection which eventually get discarded while checking the packet CRC. At higher threshold, packets gets missed in the detection function. So far, we have achieved CDMA transmission with low SINRs (0 - 5 dB) with packet error rate up to $5\%$, mostly due to the missed detection.

\section{Demo Details}
\label{sec:demo}
Our demo evaluate the scenarios of (1) multiple IoT UEs connected to an eNB and, (2) coexistence of an IoT CDMA link and an LTE link. IoT nodes are realized using USRP series X310 and/or B210. LTE transmission is enabled using openairinterface (OAI) where OAI is a PC-hosted open sourced SDR \eat{PHY layer prototyping} platform \cite{Nikaein2014_OAI}.
The LTE UE is connected to the LTE eNB using FDD mode, 5 MHz bandwidth, transmission mode 1,\eat{(SISO)} and MCS value 9 for uplink / downlink (QPSK modulation). 

\subsubsection{Data Transmission from Multiple IoT UEs}
The scenario of multiple IoT UEs connected to the single eNB poses the challenge of increased signal processing complexity at the receiver (eNB) where each UE spreads the message with separate Hadamard code and data transmission is asynchronous. Taking an example of two IoT links, we evaluate the overall implementation robustness and characterize packet-error and packet-detection rates of IoT link with respect to the SINR. The performance is evaluated under variable IoT payload (1-15 Bytes) and transmission delay between packets.\eat{{\color{blue}The scenario is also implemented and evaluated with MATLAB simulations.}} This case could further be extended for multiple IoT UEs ($>$ 2).
\eat{
The scenario of two IoT UEs connected to the single eNB poses the challenge of increased signal processing complexity at the receiver (eNB) where each UE spreads the message with separate Hadamard code and data transmission is asynchronous. This scenario is evaluated against the metrics used in the earlier scenario. This case could further be extended for multiple ($>$ 2) UEs. 
}

\subsubsection{Coexistence of IoT link and LTE}
\begin{figure}[t]
\begin{center}
\vspace{0.2in} \hspace{0.35in}
\epsfig{figure= 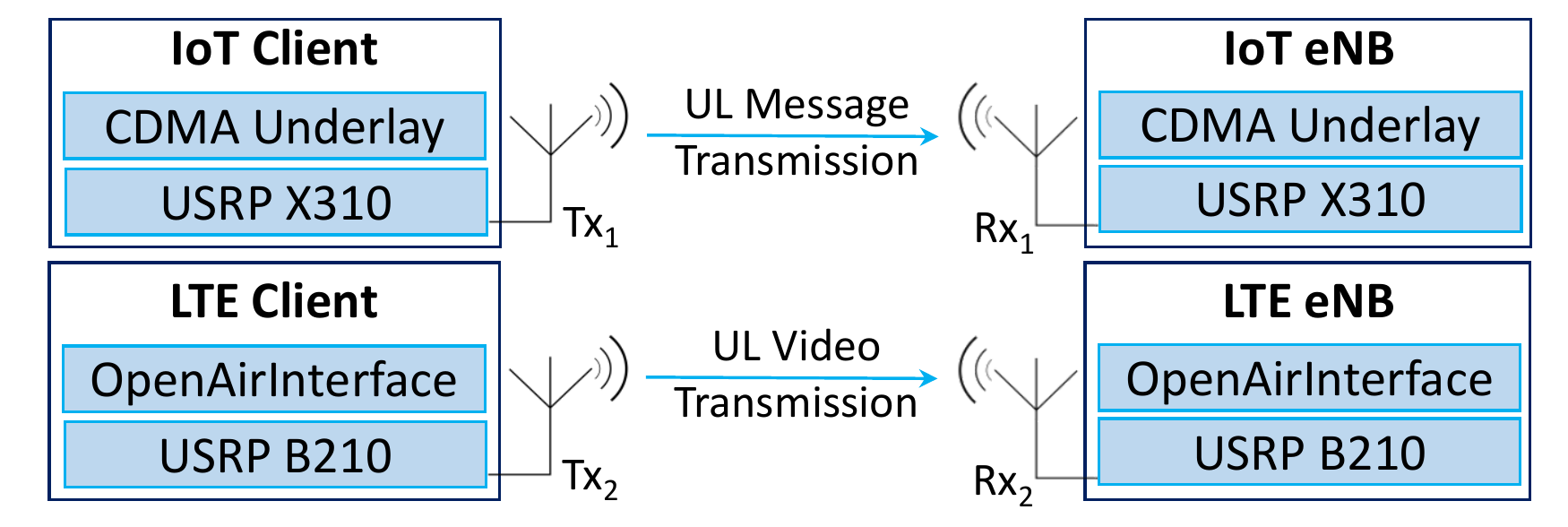,width=3.3in}
\end{center}
\caption{Setup for coexistence of IoT link and LTE using CDMA-underlay implementation, OAI and USRP}
\label{fig:IoT_LTE_coex}
\vspace{-0.6cm}
\end{figure}

Scenario shown in Fig.~\ref{fig:IoT_LTE_coex} aims to evaluate performance of the shared band operation of LTE and CDMA-underlay IoT link. 
In demo, the LTE link is used for video transmission while the CDMA channel is used for short text messages. This scenario provides insight on design details of CDMA-underlay IoT protocol considering OFDMA structure of LTE. We are currently integrating IoT eNB and OAI/LTE eNB into one unified eNB. 

\eat{
\begin{figure}[t]
\begin{center}
\vspace{0.2in} \hspace{0.35in}
\epsfig{figure= 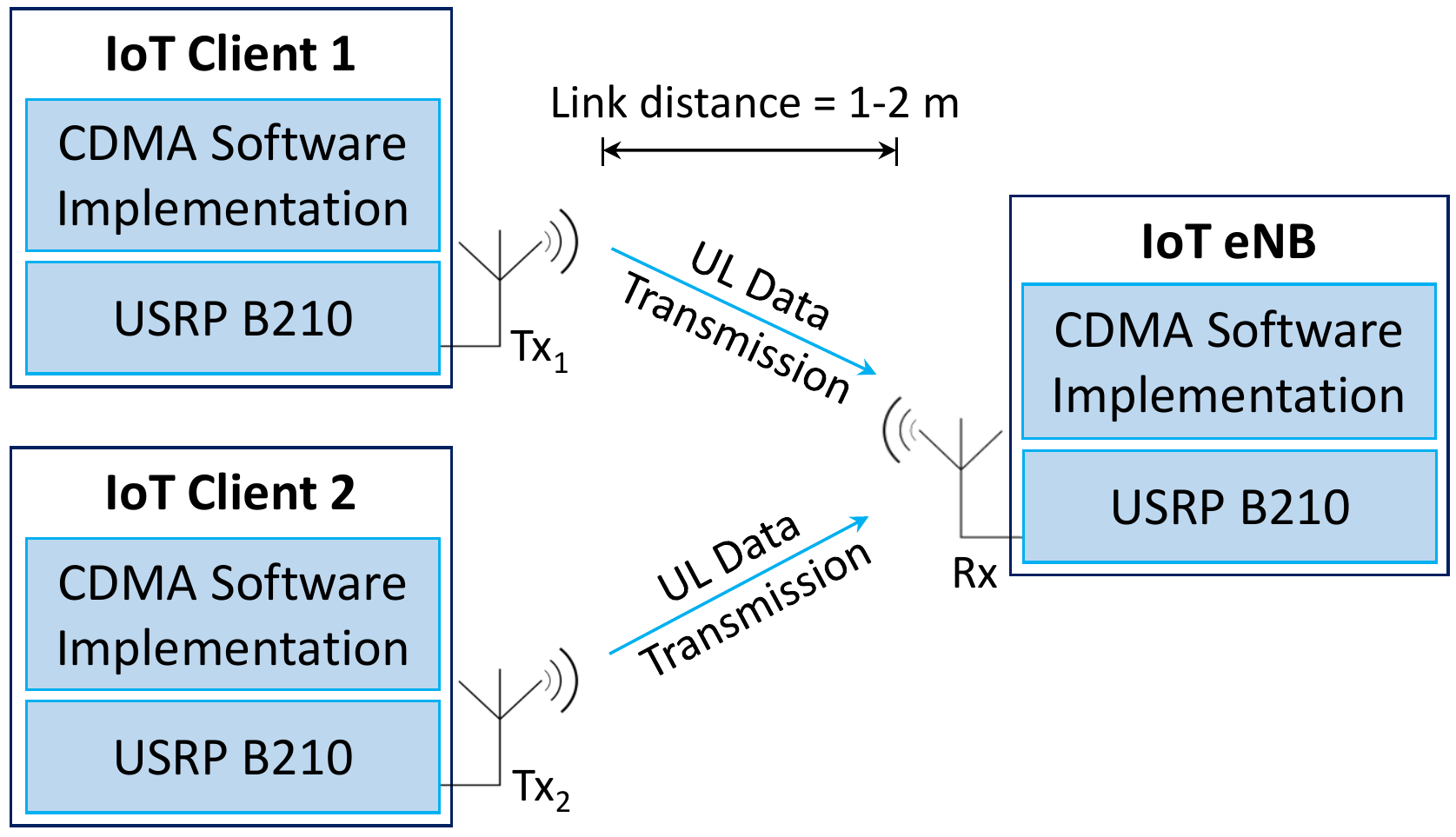,width=3.3in}
\end{center}
\caption{Setup of a single IoT link for uplink transmission using CDMA software implementation and USRP}
\label{fig:IoT_eNB_twoUEs}
\vspace{-0.6cm}
\end{figure}
}

\section{Conclusion}
With the massive deployment of IoT devices, the current 4G network will be overwhelmed by the surge in control plane signaling overhead and network latency.
This motivated us to design a CDMA-based cross-layer PHY/MAC protocol for IoT devices. The proposed IoT system coexists with in-band LTE operation and allows uplink sporadic IoT data transmission without a need of resource allocation from the LTE network.
We implemented the proposed protocol using software-defined radio platform for exemplary scenarios as a proof-of-concept.
\bibliographystyle{unsrt}
\bibliography{ref}

\begin{thebibliography}{1}

\bibitem{CiscoVNI_Feb16}
Cisco visual networking index: Global mobile data traffic forecast update, 2015
  - 2020.
\newblock White Paper.

\bibitem{Singhal2010_latencyAna}
V.~Chetlapalli D.~Singhal, M.~Kunapareddy.
\newblock {LTE}-advanced: Latency analysis for {IMT-A} evaluation.
\newblock 2010.
\newblock White Paper.

\bibitem{Mathur2016_crossLayer}
S.~Mathur, D.~Saha, and D.~Raychaudhuri.
\newblock Poster: Cross-layer {MAC/PHY} protocol to support {I}o{T} traffic in
  {5G}.
\newblock In {\em ACM Mobicom}, 2016.

\bibitem{Schaich2014_waveform}
F.~Schaich et~al.
\newblock Waveform contenders for {5G}- suitability for short packet and low
  latency transmissions.
\newblock In {\em IEEE VTC Spring}, 2014.

\bibitem{LPWA2016}
vertical m2m.
\newblock Understand {LPWA} tetchnologies ({S}igfox and {L}o{R}a), 2016.

\bibitem{Nikaein2014_OAI}
N.~Nikaein et~al.
\newblock Openairinterface: A flexible platform for 5{G} research.
\newblock {\em SIGCOMM}, 2014.

\end{thebibliography}

\end{document}